\documentclass[preprint,aps]{revtex4}

\usepackage{graphicx}

\begin{document}

\title{Electronic Origin of High Temperature Superconductivity in Single-Layer FeSe Superconductor}
\author{Defa Liu$^{1,\sharp}$, Wenhao Zhang$^{2,3,\sharp}$, Daixiang Mou$^{1,\sharp}$, Junfeng He$^{1,\sharp}$,  Yun-Bo Ou$^{3}$, Qing-Yan Wang$^{2,3}$, Zhi Li$^{3}$, Lili Wang$^{3}$, Lin Zhao$^{1}$, Shaolong He$^{1}$,  Yingying Peng$^{1}$, Xu Liu$^{1}$, Chaoyu Chen$^{1}$, Li Yu$^{1}$, Guodong  Liu$^{1}$,  Xiaoli Dong$^{1}$, Jun Zhang$^{1}$, Chuangtian Chen$^{4}$, Zuyan Xu$^{4}$, Jiangping Hu$^{3,5}$, Xi Chen$^{2}$, Xucun Ma$^{3,*}$,  Qikun Xue$^{2,*}$,   and X. J. Zhou$^{1,*}$ }

\affiliation{
\\$^{1}$National Lab for Superconductivity, Beijing National Laboratory for Condensed Matter Physics, Institute of Physics,
Chinese Academy of Sciences, Beijing 100190, China
\\$^{2}$State Key Lab of Low-Dimensional Quantum Physics, Department of Physics, Tsinghua University, Beijing
100084, China
\\$^{3}$Beijing National Laboratory for Condensed Matter Physics, Institute of Physics,
Chinese Academy of Sciences, Beijing 100190, China
\\$^{4}$Technical Institute of Physics and Chemistry, Chinese Academy of Sciences, Beijing 100190, China
\\$^{5}$Department of Physics, Purdue University, West Lafayette, Indiana 47907, USA
}
\date{February 27, 2012}
%
%




\maketitle

{\bf The latest discovery of high temperature superconductivity signature in single-layer FeSe\cite{QKXue} is significant because it is possible to break the superconducting critical temperature  ceiling (maximum T$_c$$\sim$55 K) that has been stagnant since the discovery of Fe-based superconductivity in 2008\cite{Kamihara,ZARenSm,RotterSC,MKWu11,CQJin111}. It also blows the superconductivity community by surprise because such a high T$_c$ is unexpected in FeSe system with the bulk FeSe exhibiting a T$_c$ at only 8 K at ambient pressure\cite{MKWu11} which can be enhanced to 38 K under high pressure\cite{FeSeHighP}. The T$_c$ is still unusually high even considering the newly-discovered intercalated FeSe system A$_x$Fe$_{2-y}$Se$_2$ (A=K, Cs, Rb and Tl) with a T$_c$ at 32 K at ambient pressure\cite{JGGuo,MHFang} and possible T$_c$ near 48 K under high pressure\cite{LLSun}.  Particularly interesting is that such a high temperature superconductivity occurs in a  single-layer FeSe system that is considered as a key building block of the Fe-based superconductors\cite{QKXue}. Understanding the origin of high temperature superconductivity in such a strictly two-dimensional FeSe system is crucial to understanding the superconductivity mechanism in Fe-based superconductors in particular, and providing key insights on how to achieve high temperature superconductivity in general. Here we report distinct electronic structure associated with the  single-layer FeSe superconductor. Its Fermi surface topology is different from other Fe-based superconductors; it consists only of electron pockets near the zone corner without indication of any Fermi surface around the zone center. Our observation of large and nearly isotropic superconducting gap in this strictly two-dimensional system rules out existence of node in the superconducting gap. These results have provided an unambiguous case that such a unique electronic structure is favorable for realizing high temperature superconductivity.}




Single-layer FeSe superconductor with high superconducting transition temperature is advantageous in investigating the superconductivity mechanism in a number of aspects. First, among the Fe-based superconductors discovered so far\cite{Kamihara,ZARenSm,RotterSC,MKWu11,CQJin111},  FeSe superconductor has the simplest crystal structure which consists only of FeSe layer that is the key component considered to be essential for superconductivity in Fe-based compounds\cite{MKWu11}. Second, single layer FeSe, composed of a Se-Fe-Se triple layer along the c-axis with a thickness of 5.5 $\AA$, is a strictly two-dimensional (2D) system; this renders it have 2D Fermi surface without three-dimensionality complications. For example, in typical three-dimensional bulk superconductors, in order to pin down whether there is a node in the superconducting gap, one has to exhaust all the momentum space which is usually difficult\cite{DLFeng3D}. Third, the FeSe superconductor is an ideal system where the superconducting phase is pure and well-characterized. This is in contrast with the A$_x$Fe$_{2-y}$Se$_2$ (A=K, Cs, Rb, Tl and etc.) superconducting system\cite{JGGuo,MHFang,DXReview} where there exists phase inhomogeneity and the true superconducting phase remains not clearly identified. Most importantly, the discovery of high temperature superconductivity in the single-layer FeSe\cite{QKXue} provides a platform to sort out the essential key components in realizing high temperature superconductivity in the Fe-based compounds. As the electronic structure is a fundamental requisite in dictating the physical properties and superconductivity of a material, we carried out high resolution angle-resolved photoemission spectroscopy (ARPES) on this new single-layer FeSe high temperature superconductor.

Figure 1a shows Fermi surface mapping of the single-layer FeSe thin film covering multiple Brillouin zones. In comparison, we also present Fermi surface of (Tl$_{0.58}$Rb$_{0.42}$)Fe$_{1.72}$Se$_2$ superconductor (T$_c$=32 K) (Fig. 1b)\cite{YZhang,DXMouPRL,LZhaoPRB,TQian} and (Ba$_{1.6}$K$_{0.4}$)Fe$_2$As$_2$ (T$_c$=35 K)(Fig. 1c)\cite{LZhaoCPL,HDingEPL,Borisenko,ZHasan}, as well as the Fermi surface of FeSe by band structure calculations (Fig. 1d)\cite{FeSeBandC}.  The band structure along some typical cuts are shown in Fig. 2. For the single layer FeSe superconductor,  we only observe a clear electron-like Fermi surface around M($\pi$,$\pi$)(denoted as $\gamma$ hereafter) (Fig. 1a) but do not observe any indication of Fermi surface around the $\Gamma$ (0,0) point.  The $\gamma$ Fermi surface is nearly circular with a Fermi momentum (k$_F$) of 0.25 in a unit of $\pi$/a (lattice constant a=3.90 $\AA$) which is significantly smaller than 0.35 $\pi$/a found in (Tl$_{0.58}$Rb$_{0.42}$)Fe$_{1.72}$Se$_2$ superconductor\cite{DXMouPRL}.  Considering the Fermi surface around M consists of two degenerate Fermi surface sheets (Fig. 1d), this would give an electron counting of 0.09 electrons/Fe in FeSe superconductor which is considerably smaller than that in  (Tl$_{0.58}$Rb$_{0.42}$)Fe$_{1.72}$Se$_2$ superconductor (0.18 electrons/Fe when only considering  the M point electron Fermi surface sheets). Note that this is closer to the optimal doping level in electron-doped Ba(Fe$_{1-x}$Co$_{x}$)As$_2$ (x$\sim$0.07) system\cite{NNI}.  On the other hand, the $\gamma$ band bottom in single layer FeSe lies at 60 meV below the Fermi level, slightly deeper than 50 meV in (Tl$_{0.58}$Rb$_{0.42}$)Fe$_{1.72}$Se$_2$ superconductor\cite{DXMouPRL}. The Fermi surface size and the band width together give an effective electron mass of 2.7m$_{e}$ (m$_{e}$ is free electron mass) in the single-layer FeSe and 6.1m$_e$  in (Tl$_{0.58}$Rb$_{0.42}$)Fe$_{1.72}$Se$_2$ superconductor. This indicates that in single-layer FeSe, the electrons are lighter than those in (Tl$_{0.58}$Rb$_{0.42}$)Fe$_{1.72}$Se$_2$ superconductor, possibly related to weaker electron correlation.

The observation of only electron-like Fermi surface sheets around M makes the single layer FeSe superconductor distinct from all the other Fe-based superconductors. Most of Fe-based superconductors like typical (Ba,K)Fe$_2$As$_2$  are characterized by hole-like Fermi surface sheets around the $\Gamma$ point according to band structure calculations\cite{DJSingh1111,Kuroki} and ARPES measurements(Fig. 1c)\cite{LZhaoCPL,HDingEPL,Borisenko,ZHasan}. It is also different from the Fermi surface topology of the recently discovered A$_x$Fe$_{1-y}$Se$_2$ superconductors where the Fermi surface sheets remain present around the $\Gamma$ point (Fig. 1b) although all observed Fermi surface sheets are exclusively electron-like\cite{YZhang,DXMouPRL,LZhaoPRB,TQian}. The 2D nature of the single-layer FeSe also removes any ambiguity of the possible k$_z$ complications that are usually encountered in other Fe-based systems.  We note that, compared with the band structure of (Tl$_{0.58}$Rb$_{0.42}$)Fe$_{1.72}$Se$_2$ superconductor, the band structure of single layer FeSe is not a simple rigid band shift. First, as mentioned above, the $\gamma$ band shape near M is significantly different leading to a lighter effective mass. Second, the much lower electron doping in FeSe superconductor is expected to shift the bands towards the Fermi level when comparing with the bands in (Tl$_{0.58}$Rb$_{0.42}$)Fe$_{1.72}$Se$_2$ superconductor. However, the top of the hole-like $\alpha$ band near $\Gamma$ point in FeSe superconductor (Fig. 2a1, 80 meV from the Fermi level) is similar to that in (Tl$_{0.58}$Rb$_{0.42}$)Fe$_{1.72}$Se$_2$ superconductor\cite{DXMouPRL}. The bottom of the electron-like $\gamma$ band near M in FeSe (Fig. 2a2, 60 meV from the Fermi level) is even deeper than that in (Tl$_{0.58}$Rb$_{0.42}$)Fe$_{1.72}$Se$_2$ superconductor ($\sim$50 meV from the Fermi level)\cite{DXMouPRL}, which is in opposite direction from expectation.

The identification of clear $\gamma$ Fermi surface sheet near M point makes it possible to investigate the superconducting gap in this new superconductor.  We start first by  examining the temperature dependence of the superconducting gap.  Fig. 3a shows the photoemission images along the Cut near M3 (its location shown in the bottom-left inset of Fig. 3) at different temperatures. The corresponding photoemission spectra (energy distribution curves, EDCs) on the Fermi momentum are shown in Fig. 3b where sharp peaks develop at low temperatures. To visually inspect possible gap opening and remove the effect of Fermi distribution function near the Fermi level, these original EDCs are symmetrized (see Fig. 3c), following the procedure commonly used in the study of high temperature cuprate superconductors\cite{MNorman}.   For the $\gamma$ pocket near M, there is a clear gap opening at low temperature (20 K), as indicated by a dip at the Fermi energy in the symmetrized EDCs (Fig. 3c). With increasing temperature, the dip at E$_F$ is gradually filled up and is almost invisible near 50$\sim$55 K. As it is commonly observed from ARPES on Fe-based superconductors that the gap closes at the superconducting temperature\cite{LZhaoCPL,DXMouPRL}, Fig. 3c suggests that the single layer FeSe sample we measured has a T$_c$ around (55$\pm$5K) (see ``sample temperature calibration" in Method).  This value is reproducible with an independent measurement on another sample. Note that due to technical difficulties, it is not yet possible to directly measure superconducting transition temperature on the single-layer FeSe thin film by transport or magnetic measurements\cite{QKXue}. The gap size measured at 20 K is $\sim$ 15 meV.  As a high resolution ARPES measurement on (Tl$_{0.58}$Rb$_{0.42}$)Fe$_{1.72}$Se$_2$ superconductor (T$_c$=32 K) gives a superconducting gap at $\sim$ 9.7 meV\cite{DXReview}, if we assume the same ratio between the superconducting gap size and T$_c$, 15 meV gap in the single-layer FeSe would correspond to $\sim$50 K which is within the range as estimated from the temperature dependence of the gap.


Now we come to the momentum-dependent measurements of the superconducting gap.  For this purpose we took high resolution Fermi surface mapping (energy resolution of 4 meV) of  the $\gamma$ pocket at M (upper-right inset in Fig. 4).  Fig. 4a shows photoemission spectra around the $\gamma$ Fermi surface measured in the superconducting state (T= 20 K); the corresponding symmetrized photoemission spectra are shown in Fig. 4b. We have carried out two independent measurements on two samples; the extracted superconducting gap for both samples is nearly isotropic. The sample ($\#1$) shows a superconducting gap with a size of (13$\pm$2) meV while the other sample ($\#2$) has a gap of (15$\pm$2) meV. The slight difference in the gap size is possibly due to slightly different post-annealing conditions that caused a subtle variation of the superconducting phase\cite{GapNote}.  In both samples, no indication of zero gap is observed around the $\gamma$ Fermi surface.  Since the measured single-layer FeSe is strictly two-dimensional in nature, it avoids complications from 3D Fermi surface associated with bulk materials. This establishes unambiguously that there is no node in the superconducting gap. (Fig. 4c)


Although the observed Fermi surface topology and nearly isotropic nodeless superconducting gap appear to be reminiscent to that of A$_x$Fe$_{2-y}$Se$_2$ superconductors\cite{YZhang,DXMouPRL,LZhaoPRB,TQian}, their revelation in the single-layer FeSe system with high T$_c$ provides much profound and decisive information on understanding the physics and superconductivity mechanism of Fe-based superconductors. First, in the newly-discovered A$_x$Fe$_{2-y}$Se$_2$ superconductors, because of the coexistence of many different phases,  it remains controversial on which phase is really superconducting\cite{DXReview}.  The single-layer FeSe system makes a clear-cut case that the Fe-vacancy-free FeSe phase can support high temperature superconductivity. Second, the complete absence of Fermi surface around the $\Gamma$ point removes any scattering channel between the $\Gamma$ Fermi surface sheet and M-point Fermi surface sheet. This renders it more straightforward than that in  A$_x$Fe$_{2-y}$Se$_2$ superconductors where Fermi surface sheets are still present around $\Gamma$ point albeit being electron-like. Third, because of the 2D nature of the single-layer FeSe, the present observation can make a straightforward conclusion that the superconducting gap is nearly isotropic and nodeless. In 3D A$_x$Fe$_{2-y}$Se$_2$ superconductors, such a conclusion cannot be made unambiguously before one carries out measurements for all the momentum space along all k$_z$\cite{DLFeng3D}.

The present work has important implications on the pairing mechanism in Fe-based superconductors. The electronic structure of the Fe-based compounds involves all five Fe 3d-orbitals forming multiple Fermi surface sheets: hole-like Fermi surface sheets around $\Gamma$ and electron-like ones around M\cite{DJSingh1111,Kuroki}. It has been proposed that the interband scattering between the hole-like bands near $\Gamma$ and the electron-like bands near M is responsible for electron pairing and superconductivity\cite{Kuroki,FeSCMagnetic}.  The absence of Fermi surface around $\Gamma$ point in the single-layer FeSe superconductor,  with the top of the hole-like band lying well below the Fermi level (80 meV below E$_F$, Fig. 2a1), completely rules out the electron scattering possibility across the Fermi surface sheets between the $\Gamma$ and M points. Furthermore, with the absence of Fermi surface around $\Gamma$, electrons can only scatter across the Fermi surface sheets between M points which is predicted to result in {\it d}-wave superconducting gap\cite{Kuroki,DwaveMM}.  It is argued that two Fermi surface sheets around a given M point with opposite phases can give rise to nodes around the Fermi surface although the zero-gap-line does not cross the $\gamma$ Fermi surface\cite{IMazinNode}. In this case, our identification of nodeless superconducting gap does not favor such the scheme of electron scattering across M point Fermi surface sheets as the pairing mechanism in the Fe-based superconductors.  This is in contrast to the low-T$_c$ FeSe case (T$_c$=8 K) where there is a clear indication of nodes in the superconducting gap\cite{QKScience}.  An alternative picture, such as the interaction of local Fe magnetic moment\cite{J1J2Picture} or orbital fluctuations\cite{OrbitalF}, need to be invoked to understand high temperature superconductivity in the single layer FeSe superconductor.

There are a couple of intriguing issues that merit further investigations. First, there is a disagreement between the previous tunneling measurements\cite{QKXue} and our present ARPES measurements. While two peaks at 20.1 meV and 9.0 meV are developed in the tunneling spectrum in the superconducting state of a single layer FeSe superconductor\cite{QKXue}, the present ARPES measurements do not reveal the two-gap structure in the momentum-resolved spectra\cite{GapNote}. The maximum superconducting gap ($\sim$15 meV) in our measurement is smaller than 20.1 meV but larger than 9.0 meV.  Moreover, with the identification of only one Fermi surface near M, and a nearly isotropic superconducting gap, the observation of two gaps can not be explained by multiple gaps on different Fermi surface sheets, as found in other Fe-based supercouductors like (Ba,K)Fe$_2$Se$_2$\cite{LZhaoCPL,HDingEPL}. Since tunneling is a local probe, it is highly unlikely to attribute these two peaks as due to phase separation on such a small scale.  While we are carrying out experiment to ensure that there is no change in the sample during the de-capping process of the amorphous Se protection-layer in ARPES experiment, one conjecture is whether both the FeSe thin film and the interface between FeSe and SrTiO$_3$ may become superconducting with different gap sizes. This comes to another prominent issue whether the observed superconductivity is due to FeSe itself or the interface between the FeSe and the SrTiO$_3$ substrate.  The main electronic feature of electron-like $\gamma$ Fermi surface is consistent with the band structure calculations of FeSe, and the energy gap on this particular Fermi surface closes at high temperature. Also we do not see signature of possible two-dimensional electron gas in the interface since no electron-like Fermi surface around the $\Gamma$ point is observed. These observations seem to favor the attribution of superconductivity to the FeSe layer and the tensile pressure exerted on FeSe layer due to lattice mismatch\cite{LatticeNote} between FeSe and SrTiO$_3$ may play an important role, as it has been shown that superconductivity in FeSe is rather sensitive to high pressure\cite{FeSeHighP,LLSun}. However, we cannot fully rule out interface superconductivity when considering FeSe may become superconducting due to the proximity effect and weak signal from the buried interface that may be beyond our detection. It remains to be explored why only the very first layer is superconducting while the other layers above are not superconducting\cite{QKXue}. Clearly, the answer to these questions will provide crucial information on achieving even higher T$_c$ in searching for new high temperature superconductors.

In summary, by carrying out high resolution ARPES on the single-layer FeSe high temperature superconductor, we have identified a unique Fermi surface topology that is compatible with the occurrence of high T$_c$ superconductivity in the Fe-based superconductors.  It consists of only electron-like Fermi surface near the M point.  We observed nearly isotropic superconducting gap around the Fermi surface, establishing unambiguously that this two-dimensional system exhibits  no node in the superconducting gap. These observations do not favor the electron scattering across Fermi surface sheets as the pairing mechanism of superconductivity in the Fe-based superconductors. It will shed more lights on the  nature of superconductivity in  the Fe-based superconductors and exploration of new high temperature superconductors.

{\bf\textbf{ METHODS} }

{\bf Thin film preparation and post-annealing }
The single-layer FeSe thin films were grown on SrTiO$_3$(001) substrate by the molecular beam epitaxy (MBE) method and were characterized by scanning tunneling microscope and transport measurements. The details can be found in \cite{QKXue}. In order to transfer the thin film sample from the MBE preparation chamber to a different ARPES measurement chamber, the prepared thin film is covered by an amorphous  Se capping layer.  The Se layer was desorbed by heating up the samples to 400$^\circ$C for 2 hours before ARPES measurements.

{\bf Sample temperature calibration}
Because the FeSe thin film on SrTiO$_3$ substrate is different from our usual cleaved sample, also due to their different mounting methods on the cold finger of the cryostat,  at the same temperature of the cold finger, the temperature on the FeSe thin film is higher than that for the usual cleaved sample. Therefore, the usual sample temperature sets a low limit to the temperature of the FeSe thin film. In ordr to measure the temperature of the FeSe thin film more accurately, we simulate the FeSe thin film by gluing a piece of polycrystalline gold foil on the similar SrTiO$_3$ substrate. By measuring the Fermi level of the gold from our high resolution laser photoemission, we can measure the temperature of gold on SrTiO$_3$ from the gold Fermi level width fitted by the Fermi distribution function. Since the Fermi level width contains a couple of contributions including the temperature broadening, instrumental resolution and cleanliness of gold, the estimated temperature sets an upper limit to the gold temperature which is close to the temperature of FeSe thin film on SrTiO$_3$.  We found the temperature measured from this gold reference is higher than the nominal usual sample temperature by $\sim$10 K.  The real temperature of the FeSe thin film lies then in between these two temperature limits,  nominal sample temperature and gold temperature,  with an error bar of $\pm$5 K.

{\bf High resolution ARPES methods. }

High resolution angle-resolved photoemission measurements were carried out on our lab system equipped with a Scienta R4000 electron energy analyzer\cite{GDLiu}. We use Helium discharge lamp as the light source which can provide photon energies of h$\upsilon$= 21.218 eV (Helium I).  The energy resolution was set at 10 meV for the Fermi surface mapping (Fig. 1a) and band structure measurements (Fig. 2) and at 4 meV for the superconducting gap measurements (Figs. 3 and 4). The angular resolution is $\sim$0.3 degree. The Fermi level is referenced by measuring on a clean polycrystalline gold that is electrically connected to the sample.  The sample was measured in vacuum with a base pressure better than 5$\times$10$^{-11}$ Torr.

$^{\sharp}$These people contribute equally to the present work.

$^{*}$Corresponding authors: XJZhou@aphy.iphy.ac.cn, qkxue@mail.tsinghua.edu.cn, xcma@aphy.iphy.ac.cn


\vspace{3mm}

\noindent {\bf Acknowledgement} We thank Dunghai Lee for discussions. XJZ thanks financial support from the NSFC (10734120) and the MOST of China (973 program No: 2011CB921703 and 2011CB605903). QKX and XCM thank support from the MOST of China (program No. 2009CB929400 and  No. 2012CB921702).

\vspace{3mm}

\noindent {\bf Author Contributions} X.J.Z., Q.K.X and X.C.M. proposed and designed the research. W.H.Z., Y.B.O, Q.Y.W., Z.L., L.L.W., X.C., X.C.M and Q.K.X. contributed to MBE thin film preparation. D.F.L., D.X.M., J.F.H., L.Z., S.L.H., X.L., Y.Y.P., C. Y. C., L.Y.,  G.D.L., X.L.D., J.Z., Z.Y.X., C.T.C. and X.J.Z contributed to the development and maintenance of Laser-ARPES system. D.F.L., W. H. Z.,  D.X.M., J.F.H., Y.B.O., L.Z. carried out the experiment with the assistance from S.L.H., X.L.,Y.Y.P..  J.F.H., D.X.M., L.Z. and X.J.Z. analyzed the data.  X.J.Z. wrote the paper with Q.K.X., J.F.H., D.X.M, L.Z., Y.Y.P. and J.P.H..

\newpage

\begin{figure}[tbp]
\begin{center}
\includegraphics[width=0.75\columnwidth,angle=0]{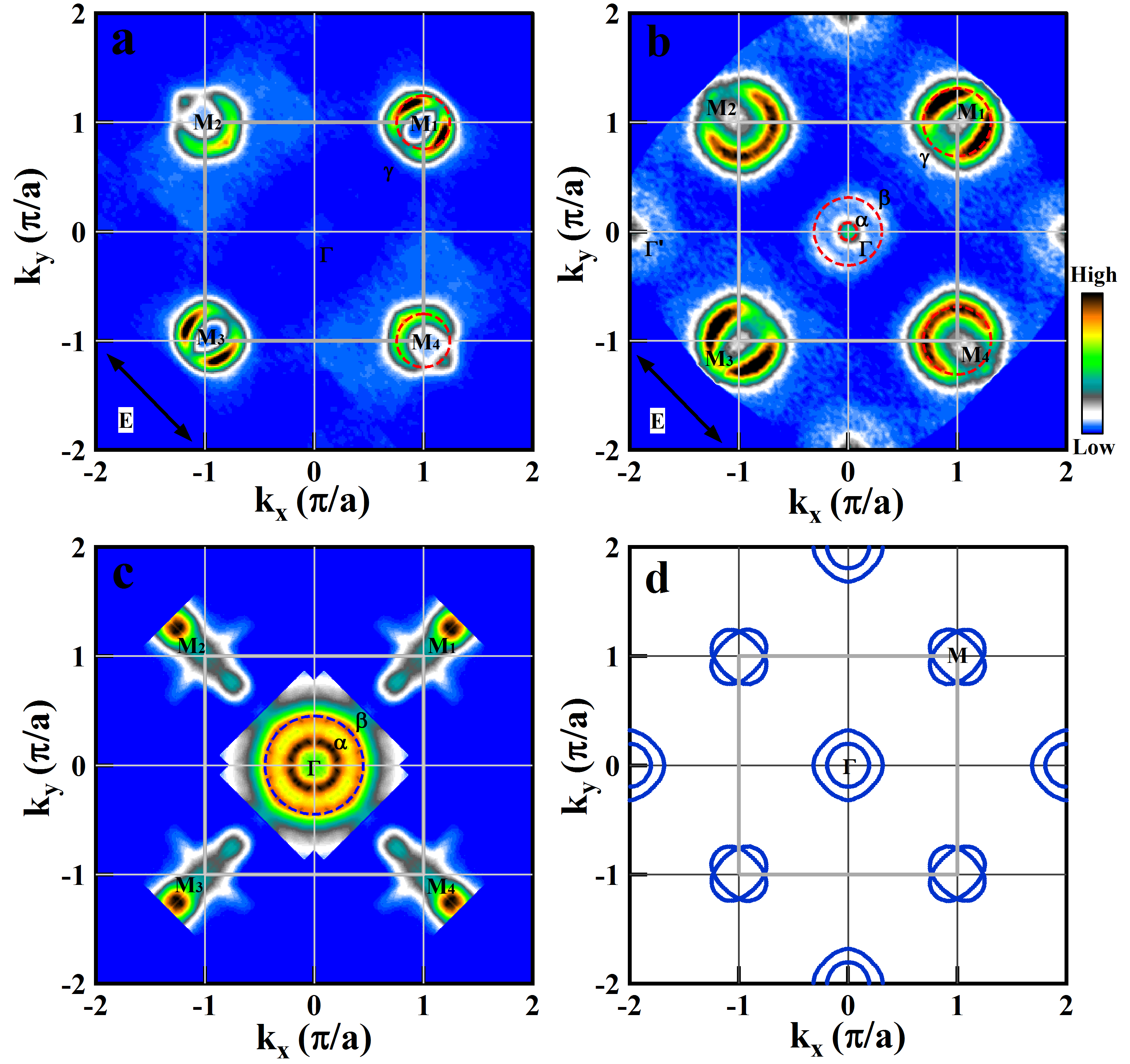}
\end{center}
\caption{Fermi surface of single layer FeSe superconductor in comparison with other Fe-based superconductors and band structure calculations.  (a). Fermi surface mapping of single layer FeSe measured at 20 K which consists only of the electron-like Fermi surface sheet ($\gamma$) around M($\pi$,$\pi$).  For convenience, the four equivalent M points are labeled as M1($\pi$,$\pi$), M2(-$\pi$,$\pi$), M3(-$\pi$,-$\pi$) and M4($\pi$,-$\pi$).  (b). Fermi surface mapping of (Tl$_{0.58}$Rb$_{0.42}$)Fe$_{1.72}$Se$_2$ superconductor (T$_c$=32 K) which consists of electron-like Fermi surface sheet ($\gamma$) around M and also electron-like Fermi surface  sheets ($\alpha$ and $\beta$) around $\Gamma$(0,0)\cite{DXMouPRL}. (c). Fermi surface mapping of (Ba$_{1.6}$K$_{0.4}$)Fe$_2$As$_2$ superconductor (T$_c$=35 K) which consists of hole-like Fermi surface sheets ($\alpha$ and $\beta$) around $\Gamma$ and complex Fermi surface near M region\cite{LZhaoCPL}. Note that the absence of spectral weight near the 2nd zone centers is because the measured region did not cover this area.  (d). Fermi surface of $\beta$-FeSe by band structure calculations for k$_z$=0 (blue thick lines) which consists of hole-like Fermi surface sheets around $\Gamma$ and two electron-like Fermi surface sheets around M\cite{FeSeBandC}.
}
\end{figure}

\begin{figure*}[tbp]
\begin{center}
\includegraphics[width=1.0\columnwidth,angle=0]{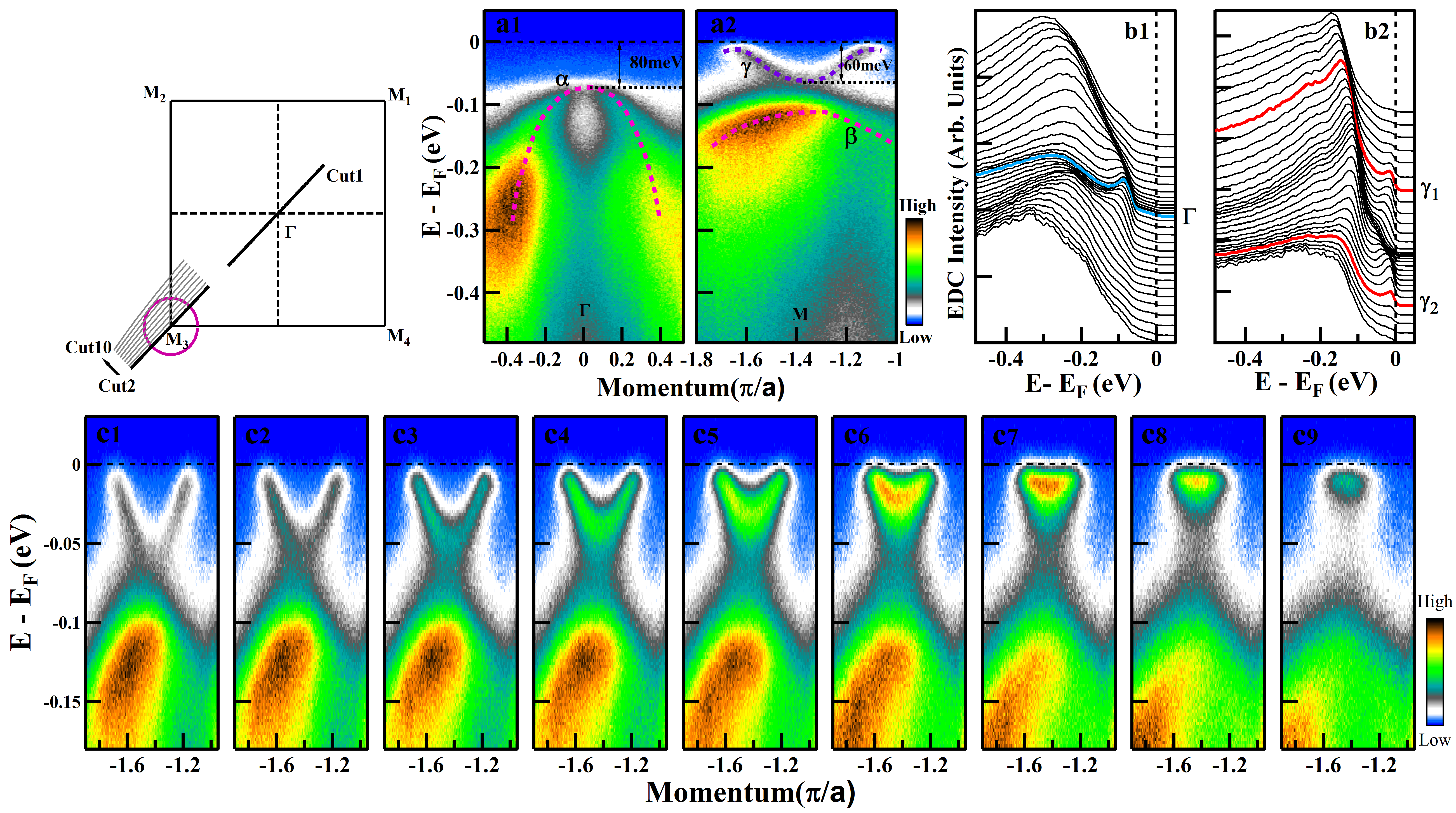}
\end{center}
\caption{Band structure and photoemission spectra of single layer FeAs superconductor. The upper-left inset shows the location of the momentum cuts along some high symmetry lines and near the M3 region.  (a). Band structure along the Cut 1 crossing the $\Gamma$ point (a1) and along the Cut 2 crossing the M3 point (a2).   (b). Photoemission spectra (energy distribution curves, EDCs) corresponding to image (a1) for the Cut 1 (b1) and corresponding to image (a2) for the Cut 2.  (c). Detailed band structure evolution near the M3 region from the Cut 2 (C1) to the Cut 10 (C9).
}
\end{figure*}

\begin{figure}[tbp]
\begin{center}
\includegraphics[width=1.0\columnwidth,angle=0]{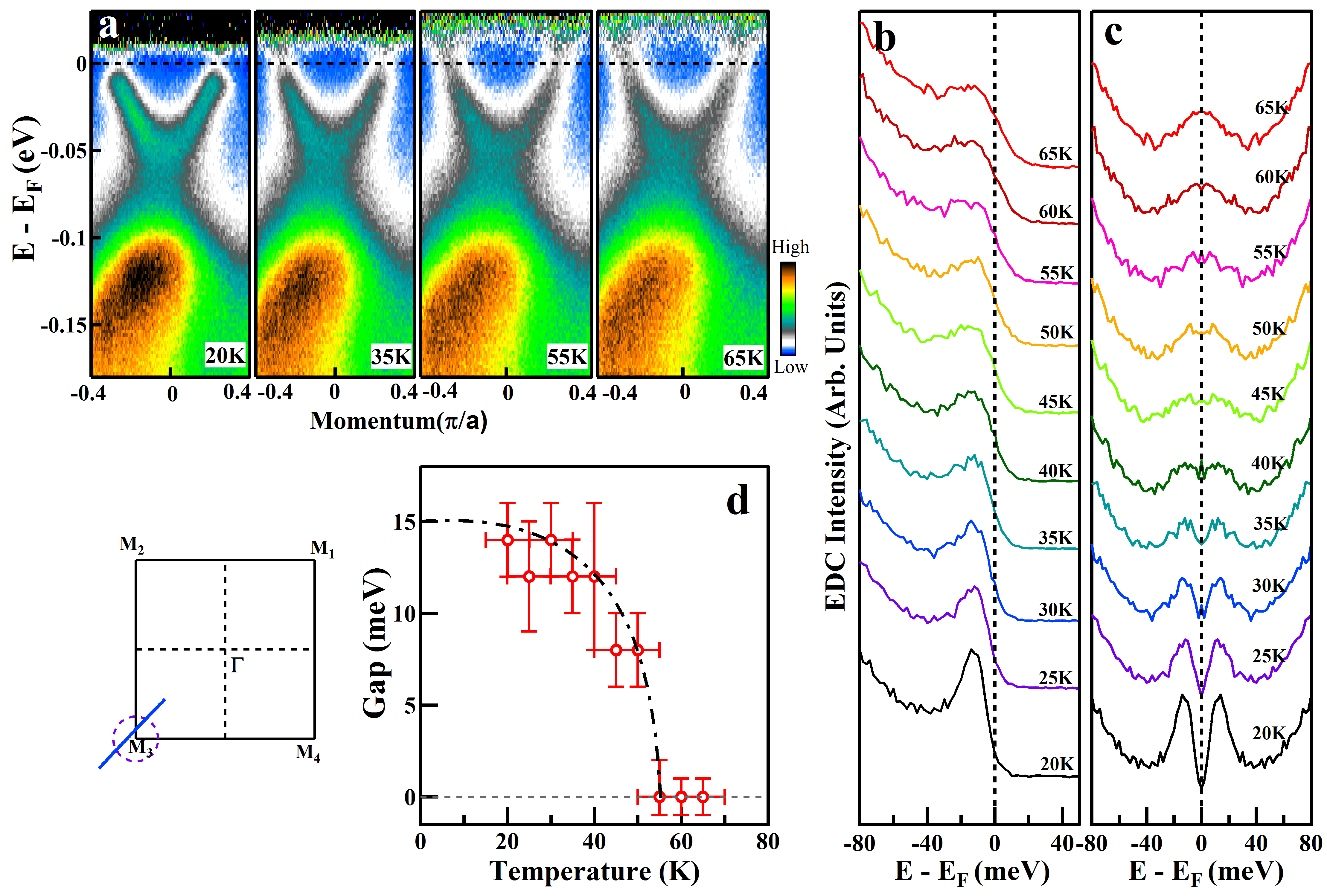}
\end{center}
\caption{Temperature dependence of energy bands and superconducting gap of single layer FeSe superconductor near M points. (a). Typical photoemission images along the Cut near the M3 point (bottom-left inset) measured at different temperatures. The temperature labeled here is a usual nominal sample temperature (see ``sample temperature calibration" in Method).  Each image is divided by the corresponding Fermi distribution functions in order to highlight opening or closing of an energy gap.  (b). Photoemission spectra at the Fermi crossings of $\gamma$ Fermi surface  and their corresponding symmetrized spectra (c) measured at different temperatures. (d). Temperature dependence of the superconducting gap. The dashed line is a BCS gap form which gives a gap size at zero temperature of 15 meV.
}
\end{figure}

\begin{figure}[tbp]
\begin{center}
\includegraphics[width=1.0\columnwidth,angle=0]{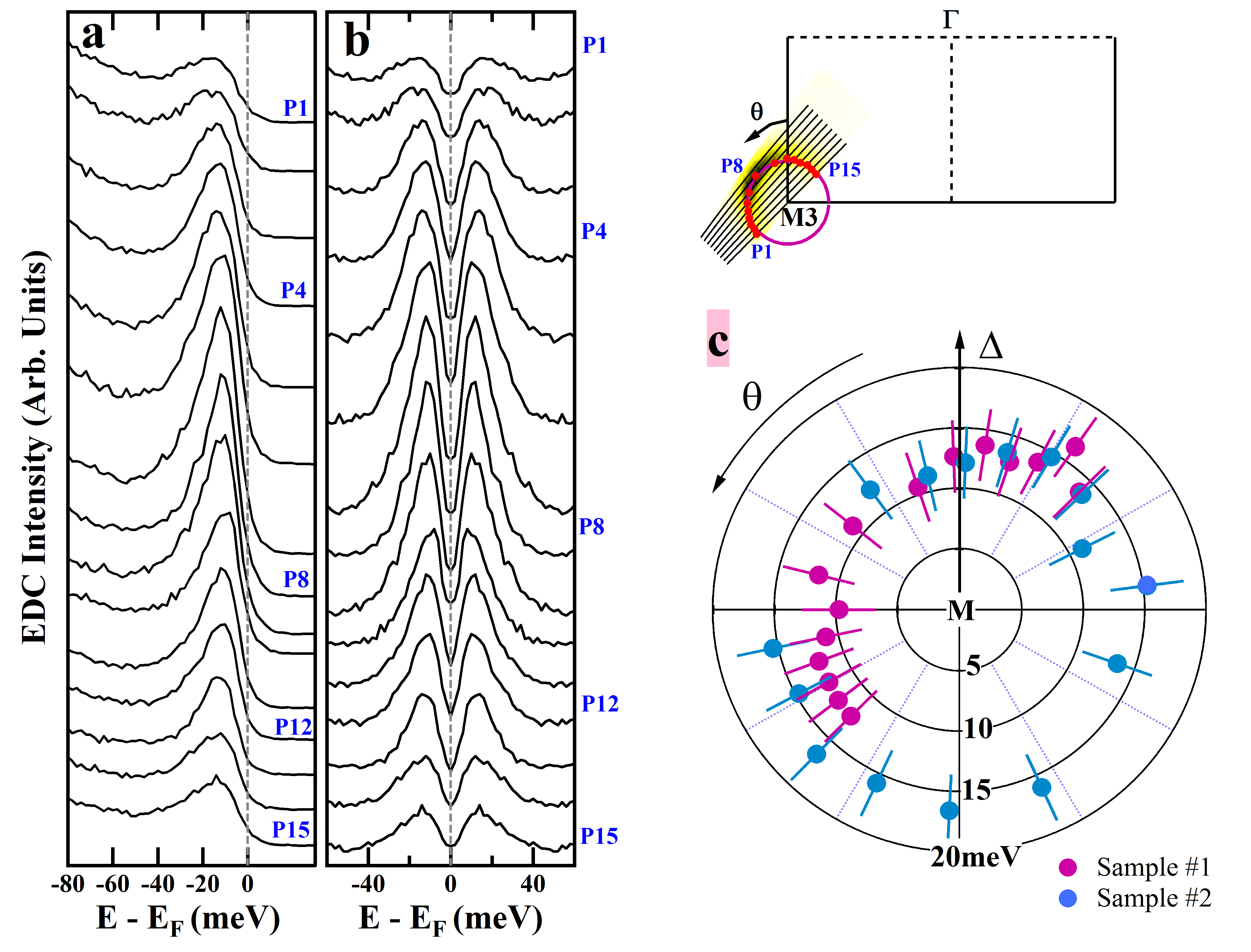}
\end{center}
\caption{Momentum dependent superconducting gap of single layer FeSe superconductor (Sample $\#1$) along the $\gamma$ Fermi surface measured at T=20 K. The Fermi surface mapping near M3 and the corresponding Fermi crossings are plotted in the top-right corner.  (a). EDCs along the $\gamma$ Fermi surface and (b) their corresponding symmetrized EDCs.  (c). Momentum dependence of the superconducting gap, obtained by picking the peak position in the symmetrized EDCs, along the $\gamma$ Fermi surface. Two superconducting samples are independently measured and their superconducting gaps are both shown with the pink solid circles representing the gap of Sample $\#$1 and the blue solid circles representing the gap of Sample $\#$2.
}
\end{figure}


\begin{thebibliography}{99}
\bibitem{GapNote}We note that the superconductivity in single-layer FeSe is sensitive to the preparartion and post-annealing conditions. It is under investigation whether the superconductivity of the Se-capped and later de-caped sample measured by ARPES is identical to that prepared and meaured in situ by tunneling.
\bibitem{LatticeNote}The lattice constant for bulk FeSe is a=3.76 $\AA$\cite{MKWu11},  for A$_x$Fe$_{2-y}$Se$_2$ is a=3.892 $\AA$, and for SrTiO$_3$ is a=3.905 $\AA$. This means when FeSe is epitaxially grown on SrTiO$_3$(001) substrate, the apparent lattice mismatch between FeSe and SrTiO$_3$  will exert a strong tensile strain on the FeSe layer. On the other hand, such a mismatch is much reduced in A$_x$Fe$_{2-y}$Se$_2$.
\end{thebibliography}
\end{document}